\documentclass[superscriptaddress,amssymb,amsmath,nofootinbib,twocolumn,aps,prd,longbibliography]{revtex4-1}
\usepackage[utf8]{inputenc}
\usepackage{mathtools}
\usepackage{hyperref}
\usepackage{graphicx}
\usepackage{slashed}
\hypersetup{colorlinks=true, urlcolor=blue,linkcolor=blue, citecolor=blue}
\usepackage{longtable}

\DeclarePairedDelimiterX\MeijerM[3]{\lparen}{\rparen}%
{\,#3\delimsize\vert\begin{smallmatrix}#1 \\ #2\end{smallmatrix}}
\newcommand\MeijerG[7]{%
	\text{G}^{#1,#2}_{#3,#4}\MeijerM*{#5}{#6}{#7}}

\begin{document}
\title{Renormalons beyond the Borel plane}
\author{E. Cavalcanti}
\email[]{erich@cbpf.br}
\affiliation{Centro Brasileiro de Pesquisas F\'{\i}sicas/MCTI, 22290-180 Rio de Janeiro, RJ, Brazil}

\begin{abstract}
The renormalon singularities are a known source of the divergent behavior of asymptotic perturbative series from field theoretical models. These singularities live in the Borel plane and are responsible for ambiguities in the physical plane. We propose that field theories can have renormalons beyond the usual first Borel plane. We show an example with a scalar field theory where, considering a chain of cat's eyes diagrams, the model develops a Gevrey-3 asymptotic series.
\end{abstract}

\maketitle
    

\section{Introduction}
\label{sec:intro}

The renormalon phenomenon is known since the `70s; it was initially proposed by 't Hooft, who noted it as one source of divergence for the perturbative asymptotic series of quantum field theories. The renormalon shows up as a singularity in the Borel plane. Initially, the exact definition of the renormalon singularity was a matter of discussion, with some debate whether it was a simple pole, a $n$-th order pole, or rather a branch cut on the Borel plane~\cite{Beneke:1998ui}. Although new perspectives have draw attention in the past two decades~\cite{Loewe:1999kw,Kneur:2001dd,Guralnik:2007rx,Unsal:2008ch,Argyres:2012ka,Argyres:2012vv,Dunne:2012ae,Dunne:2012zk,Poppitz:2012nz,Cherman:2013yfa,Shifman:2013uka,Dabrowski:2013kba,Cherman:2014ofa,Shifman:2014fra,Anber:2014sda,Sulejmanpasic:2016fwr,Sulejmanpasic:2016llc,Dunne:2016nmc,Buividovich:2017jea,Caprini:2017ikn,Aniceto:2018bis,Cherman:2018mya,Aniceto:2018bis,Cherman:2018mya,Maiezza:2018ags,Cavalcanti:2018thz,Fujimori:2018kqp,Marino:2019wra,Marino:2019eym,Marino:2019fvu,Correa:2019xvw,Ishikawa:2019oga,Ishikawa:2019tnw,Ashie:2019cmy,Correa:2019xvw,Borinsky:2020vae,Cavalcanti:2020uib,Dondi:2020qfj,Ashie:2020bvw,Borinsky:2020vae,Unsal:2020yeh,Costin:2020hwg}, the seminal report from Beneke, see ref.~\cite{Beneke:1998ui}, is still a useful reference on state of the art regarding the formal discussion.

Nowadays, the renormalon problem is understood as a component of the broader context of the resurgence program~\cite{Anber:2014sda,Dunne:2016nmc,Maiezza:2018ags,Marino:2019wra,Marino:2019eym,Marino:2019fvu,Ishikawa:2019oga,Dondi:2020qfj}. The singularity is an indication that the original perturbative series requires a transseries contribution. Also, the renormalon singularity, which arose as a formal aspect from quantum field theory, is seen as a useful guide to phenomenological predictions~\cite{Cvetic:2019jmu,Hadjimichef:2019vbb,Takaura:2019vak,Hayashi:2019mlb,Corcella:2019tgt,Maiezza:2019dht,Nason:2019gff,Cvetic:2018qxs,Ortega:2018oae,Kataev:2018mob,Mateu:2018zym,Boito:2018dmm,Boito:2018rwt,FerrarioRavasio:2018ubr,Brambilla:2018tyu,Braun:2018brg,Kataev:2018fvx,Bell:2018gce,Takaura:2018lpw,Takaura:2018vcy,Suzuki:2018vfs,DelDebbio:2018ftu,Kataev:2018gle,Peset:2018ria,Bazavov:2018omf,Brambilla:2017hcq,Nason:2017cxd,Ortega:2017lyb,Mateu:2017hlz,Hoang:2017btd,Ahmadov:2017fcf,Steinhauser:2016xkf,Jamin:2016ihy,Ayala:2016sdn,Mishima:2016vna}.

It is usual in the literature to consider a large-$N$ expansion when one is interested in exploring the existence of renormalons inside the theory. That occurs because, in this limit, the relevant contributions are planar diagrams, more commonly chains of bubble diagrams. For example, for quantum chromodynamics (QCD) this produces the usual contribution to the Adler function. However, the large-$N$ expansion hides other singularities, as the instanton problem, because the poles get far away from the origin in this approximation. Due to historical reasons -- as pointed by ref.~\cite{Beneke:1998ui} -- it is very common to relate renormalons and bubble-chain diagrams, although the renormalon singularity might appear at different subsets of Feynman diagrams. Once we consider a quantum field theory in general (outside the large-$N$ approximation), we have no guarantee that the usual planar diagrams are the most relevant ones, and other scenarios might appear. 

Another common relationship is that the renormalon singularities live in the Borel plane. To our knowledge, every report so far is still considering that renormalons are singularities that lives in the  first Borel plane and produce a Gevrey-1 asymptotic series~\cite{Marino:2020dgc,Marino:2019fvu,Marino:2019eym,Ishikawa:2019oga,Ishikawa:2019tnw,Cavalcanti:2018thz,Cavalcanti:2020uib,Ashie:2020bvw}. Here we propose a new perspective, where we suppose a generic chain and show that new kinds of singularities occur that might live in a $q$-Borel plane instead of the typical Gevrey-1 case of the Borel plane. Therefore, we propose an extension on the definition of the renormalon singularity, asserting that they live in a Gevrey-$q$ plane, thus extending the Gevrey-order of the perturbative series.

We aim to identify scenarios where a renormalon singularity might occur. Notice that we do not intend to discuss at which order in some perturbative expansions these contributions appear. The relevant point here is that the contributions considered are subsets of the full perturbative expansion and shall appear. Following this chain of thought, we also do not apply any procedure to circumvent the renormalon problem and give the nonperturbative solution, as we are mostly interested in studying the existence or not of renormalons singularities.

As our main interest lies in the formal aspect of quantum field theory, it suffices to consider as a toy model a massless scalar field theory. This model is known to possess ultraviolet (UV) renormalons produced by a sum of bubble-chain diagrams, and -- more importantly -- we show that this model reveals one novel scenario with cat's eyes chain diagrams. The sum over these chains produces a Gevrey-3 series, therefore extending the definition of the renormalon singularity beyond the usual Borel plane.


\section{$q$-Borel series}

In many scenarios, perturbative series diverge and only make sense as an asymptotic series. A rather simple example of divergence occurs when the perturbative series has a factorial growth, as
\begin{equation}
A \sim \sum_{k=0}^\infty (k!)^q (s)^k g^k,
\end{equation}
\noindent where $g$ is the expansion parameter (the coupling constant), $q$ is the power of the factorial growth, and $s$ is the sign ($s=\pm 1$) of the coefficients.

A well-stablished procedure is to consider an extension of the Borel sum~\cite{Bender:2013book}. The usual Borel sum occurs for $q=1$ and relates to the conventional asymptotic power series (in the sense of Poincaré). In this extension we consider $q$ factorials to control the growth, which is well established in the studies of $q$-summability and Gevrey-$q$ series. First we transport the usual sum to a summation in the $q$-Borel plane
\begin{equation}
A \rightarrow B_q[A](u) \sim \sum_{k=0}^\infty (su)^k = \frac{1}{1-s u}.
\end{equation}
Then, if this new infinite sum is summable in the $q$-Borel plane we proceed to implement a $q$-Borel inverse transform
\begin{equation}
\widetilde{A}[g] = \int_0^\infty du_1 \ldots du_q e^{-(u_1+\cdots+u_q)} B_q[A](g u_1 \ldots u_q).
\end{equation}

In the following, we denote the physical plane as 0-Borel (Gevrey-0), the usual Borel plane as 1-Borel (Gevrey-1), and so on. 

Let us consider, as an example, $q=1$ and $q=2$ in the most simple scenario. For $s=-1$, 
\begin{align*}
-\frac{e^{\frac{1}{g}}}{g}\text{Ei}\left(-\frac{1}{g}\right),&\quad q=1;\\
\frac{\pi}{g}  \MeijerG{1}{1}{3}{1}{0, \frac{1}{2} }{0,0,0,\frac{1}{2}}{-\frac{1}{g}}
,&\quad q=2;
\end{align*}
\noindent where $\text{Ei}$ is the exponential integral function and $\text{G}^{m n}_{p q}$ is the Meijer-$G$ function. Therefore, we can relate the asymptotic sum to a well-defined function through $q$-summability.

On the other hand, if we take $s=1$, we find a pole in the positive real axis of the $q$-Borel plane. This singularity introduces an ambiguity due to the choice of the integration path.

Few known physical models produce a Gevrey-$q$ series and discuss $q$-summability. For example, a scenario with Gevrey-2 is the sextic anharmonic oscillator, while for Gevrey-3 there is the octic anharmonic oscillator~\cite{Shalaby:2020gbk, Weniger:1996aaa}.

In sec.~\ref{sec:renormalons}, we propose with some degree of generality that there can be Gevrey-$q$ series in field theoretical models. This is shown both for UV and IR renormalons. In sec.~\ref{sec:renormalons:eyecat}, we take a toy model and show explicitly the existence of a Gevrey-3 series when considering the subset of cat's eyes chain diagrams.

\section{Renormalons from chains}
\label{sec:renormalons}

Renormalons are singularities that arise due to low/large momenta of integration (IR/UV renormalons) in the summing of a particular subset of Feynman diagrams. The usual understanding is that they live in the first Borel plane. In this section, we argue that they can occur in the $q$-Borel plane, meaning that we are not considering the Gevrey-1 asymptotic series as usual, but that the asymptotic series can be of the Gevrey-$q$ type.

In the following we consider the sum over a subset of Feynman diagrams that produce a ``chain", $R_k(\alpha)$ (fig.~\ref{fig:eyecatchaink}). The chain is built by the successive introduction of ``chain-links", $g(\ell)$ (fig.~\ref{fig:eyecatlink}). The chain link is some simple structure as a bubble diagram, a sunset diagram, a cat's eye diagram or other possibilities. Formally, we can write the sum over all chain diagrams of some particular chain link as
\begin{equation}
R(\alpha) = \sum_k \int d\ell f(\ell) \left[\alpha g(\ell)\right]^k,
\end{equation}
\noindent where $\alpha$ is related to the coupling of the theory, and $f(\ell)$ contains all additional contribution for the chain diagram. The momentum $\ell$ is integrated over the whole chain.

This representation is very general and contains all possibilities for a one-chain scenario. We remark that we do not deal here with the multi-chain scenario.

To determine the structure of $R(\alpha)$ we are mainly interested in the asymptotic behavior with respect fo the internal momentum $\ell$ (this translates to consider the large $k$ behavior. The behavior of both $f(\ell)$ and $g(\ell)$ depends on the particular scenario under interest. However, it is known from the expression of Feynman amplitudes in the complete Mellin representation~\cite{Bergere:1977wz,Smirnov:1994tg,Linhares:2007pt}, that the asymptotic expansion with respect to the external momentum behaves as 
\begin{equation}
I_G \sim \ell^p \ln^q \ell, 
\end{equation}
\noindent where $p \in \mathbb{Z}$, $q \in \mathbb{Z}^+$.

Therefore, we can say without loss of generality\footnote{In a recent article, ref.~\cite{Marino:2019fvu} considered a scenario that behaves as $\ln x/(x+1)$. That produces renormalon poles in scenarios prohibited here. However, this expression is ``unstable", a small perturbation in the asymptotic approximation ($\ln x \rightarrow \ln 1+x$) destroys the renormalon pole.} that both $f(\ell)$ and $g(\ell)$ behaves asymptotically as a transmonomial\footnote{The transmonomial is an extension of the usual monomial structure and is related to the definition of a transseries.} $\ell^p \ln^q \ell$.

\subsection{UV renormalons}
\label{sec:renormalons:UV}

We have to split the investigation into the IR and UV scales, as the asympototic behavior is different in each case. First, let us consider the UV scale
\begin{subequations}
	\begin{align}
	f(\ell) &= \ell^{-a} \ln^b \ell, \quad
	g(\ell) = \ell^{-c} \ln^d \ell,\\
	R_{UV}(\alpha) &\sim \sum_k \int_1^\infty d\ell f(\ell) \left[\alpha g(\ell)\right]^k \nonumber\\&= \sum_k \alpha^k \int_1^\infty d\ell\; \ell^{-a-ck} \ln^{b+dk} \ell,
	\label{eq:Ruv}
	\end{align}
\end{subequations}
\noindent here $a\ge 0, c\ge 0$ such that the integration does not require any new subtraction and is well behaved for $\ell \rightarrow \infty$.

Making the change of variables $\ell = e^t$ and identifying the gamma function,
\begin{equation}
R_{UV}(\alpha) \sim \sum_k \alpha^k \frac{\Gamma(dk+b+1)}{(a+ck-1)^{dk+b+1}}.
\end{equation}
\noindent We obtain the large $k$ behavior using the Stirling approximation for the gamma function,
\begin{equation*}
R_{UV}(\alpha) \sim 
\sum_k \alpha^k \left(\frac{dk+b}{e(a+ck-1)}\right)^{dk+b} \frac{\sqrt{2\pi(dk+b)}}{a+ck-1}.
\end{equation*}

In the scenario with $c\neq 0$ the $k^k$ behaviour related to the factorial growth is cancelled out and the function is well defined,
\begin{multline*}
R_{UV}(\alpha) \sim \sum_k \alpha^k \left(\frac{d}{ec}\right)^{dk+b} \frac{\sqrt{2\pi d}}{c \sqrt{k}} \\
= \frac{\sqrt{2\pi d}}{c} \left(\frac{d}{ec}\right)^{b} \text{Li}_{1/2} \left( \left(\frac{d}{ec}\right)^{d}\alpha \right).
\end{multline*}

The relevant scenario occurs only if $c=0$ and $d>1$, at large $k$ we obtain
\begin{multline}
R_{UV}(\alpha) \sim 
\frac{1}{(2\pi)^{\frac{d-1}{2}}}  \frac{d^{b+\frac{1}{2}}}{(a-1)^{b+1} e^b}  \times\\\sum_k \frac{k^b k!^d}{(2\pi k)^{\frac{d-1}{2}}} \left[\alpha\left(\frac{d}{a-1}\right)^{d} \right]^k.
\end{multline}

\noindent This expression reveals that in the UV regime the sum over the set of chain diagrams behaves as a Gevrey-$d$ series. The special case of $d=1$ produces the usual Grevrey-1 scenario of the Borel sum. In this approximation the sum of chains is $d$-Borel summable and produces a polylogarithm of order $(d-1)/2-b$,

\begin{multline}
R_{UV}(\alpha) \rightarrow B_d[R_{UV}](u) \sim 
\frac{1}{(2\pi)^{\frac{d-1}{2}}}  \frac{d^{b+\frac{1}{2}}}{(a-1)^{b+1} e^b} \times\\\sum_k 
\left[\left(\frac{d}{a-1}\right)^{d} \text{sign}(\alpha)u\right]^k \frac{k^b}{k^{\frac{d-1}{2}}}
\\  =\frac{d^{b+\frac{1}{2}}\text{Li}_{\frac{d-1}{2}-b} \left[\left(\frac{d}{a-1}\right)^{d} \text{sign}(\alpha)u\right]}{(2\pi)^{\frac{d-1}{2}}(a-1)^{b+1} e^b} .
\end{multline}

The polylogarithm function of order $\nu$, $\text{Li}_\nu(x)$, has a branch point at $x=1$ if $\nu>0$, or a pole of order $1-\nu$ if $\nu\le 0$. Therefore, if $\frac{d-1}{2}-b > 0$ we have a branch point at $u= u_0 = \text{sign}(\alpha) \left(\frac{a-1}{d}\right)^d$. And, if $\frac{d-1}{2}-b \le 0$ this point is a pole of order $1+b + \frac{1-d}{2}$.

For example, at perturbative QCD, see ref.~\cite{Beneke:1998ui}, $\alpha = -|\alpha|$ and in UV region (for the chain of fermionic bubbles at $1/N_f$, where $N_f$ is the number of flavors) we have $a=2,b=1,c=0,d=1$ This means that $\nu = 1$ and we obtain a Gevrey-1 UV renormalon located at $u=-1$ as a double pole (as expected).

\subsection{IR renormalons}
\label{sec:renormalons:IR}

To take into account the IR scale we use a slightly different choice of $f(\ell)$ and $g(\ell)$,
\begin{subequations}
	\begin{align}
	f(\ell) &= \ell^{a} \ln^b \ell, \quad 
	g(\ell) = \ell^{c} \ln^d \ell,\\
	R_{IR}(\alpha) &\sim \sum_k \int_0^1 d\ell\; f(\ell) \left[\alpha g(\ell)\right]^k \nonumber\\&= \sum_k \alpha^k \int_0^1 d\ell\; \ell^{a+ck} \ln^{b+dk} \ell.
	\end{align}
\end{subequations}

Note that the power dependence with $\ell$ is in the numerator instead of the denominator as in the UV scale. This choice is to guarantee that $R_{IR}$ is well behaved in the IR scale and does not require any new subtraction.

We can employ the substitution $\ell = 1/v$,  
\begin{multline*}
R_{IR}(\alpha) \sim \sum_k \alpha^k \int_1^\infty dv\; \frac{1}{v^2} v^{-a-ck} (-\ln v)^{b+dk} 
\\= \sum_k (-1)^b\left((-1)^d\alpha\right)^k \int_1^\infty dv\; \frac{1}{v^2} v^{-a-ck} (\ln v)^{b+dk},
\end{multline*}

\noindent  the integral assumes the same form as $R_{UV}$, see eq.~\eqref{eq:Ruv}, with the little shift $a\rightarrow a+2$.  Following the same steps as before we obtain that -- requiring $c=0$ and $d>1$ -- in the IR regime the sum of chains also behaves as a Gevrey-$d$ series. At large $k$  it produces
\begin{multline}
R_{IR}(\alpha) \sim
\frac{1}{(2\pi)^{\frac{d-1}{2}}}  \frac{(-1)^b d^{b+\frac{1}{2}}}{(a+1)^{b+1} e^b} \times\\ \sum_k \frac{k^b k!^d}{(2\pi k)^{\frac{d-1}{2}}} \left[\alpha\left(-\frac{d}{a+1}\right)^{d} \right]^k,
\end{multline}

\noindent which is $d$-Borel summable, producing a polylogarithm of order $(d-1)/2-b$  with a singularity at
\begin{equation*}
u = \left(-\text{sign}(\alpha)\frac{a+1}{d}\right)^d,
\end{equation*}
\noindent once again this singularity is a pole of order $1+b+(1-d)/2$ in the scenario where $(d-1)/2-b\le 0$, or a branch point if $(d-1)/2-b > 0$.

At perturbative QCD, see ref.~\cite{Beneke:1998ui}, $\alpha = -|\alpha|$ and in IR region we have $a=1,b=0,c=0,d=1$ for the chain of fermionic bubbles at $1/N_f$, where $N_f$ is the number of flavors. Also $\nu = 1$, so $u=2$ is the first IR renormalon and is a simple pole (as expected).

\subsection{Discussion}
\label{sec:renormalons:discussion}

In the last sections, we indicate
that the existence of renormalons is directly related to the logarithm behavior of the chain links. This relationship is commonly assumed, although not explicitly formulated. Therefore we propose to conjecture that:
\begin{itemize}
	\item the sum of a set of chains \textit{can have} a renormalon \textit{only} if the chain-link behaves as $\ln \ell$.
\end{itemize}

About the renormalon singularity, we also claim that:
\begin{itemize}
	\item The renormalon ``lives" in the $d$-Borel plane, where $\ln^d \ell$ is the behavior of each ``insertion"/chain-link.
	\item The renormalon is a pole of order $1 + b + (1-d)/2$, where $b$ is the logarithm contribution, $\ln^b \ell$, outside the insertions. It can also be a branch point if $b + (1-d)/2<0$.
	\item The location of the UV/IR pole depends only on $a$ and $d$, where $a$ is the power contribution ($\ell^a$ for IR renormalons, $\ell^{-a}$ for UV renormalons) outside the insertions. 
\end{itemize}

With this, it becomes clear that renormalons do not need to be singularities in the first Borel plane; this is just a particular case.

In what follows, we exhibit a simple example where a Gevrey-3 renormalon appears. 

\section{cat's eyes chain}
\label{sec:renormalons:eyecat}

Let us consider a flavorless and massless scalar field theory with quartic interaction $\lambda\phi^4/4!$. We propose to build a chain diagram where each chain link is given by a cat's eye diagram. Each insertion is 
\begin{equation}
G(\ell) = (-g)^3 I(\ell),
\end{equation}

\begin{figure}
	\centering
	\includegraphics[width=0.5\linewidth]{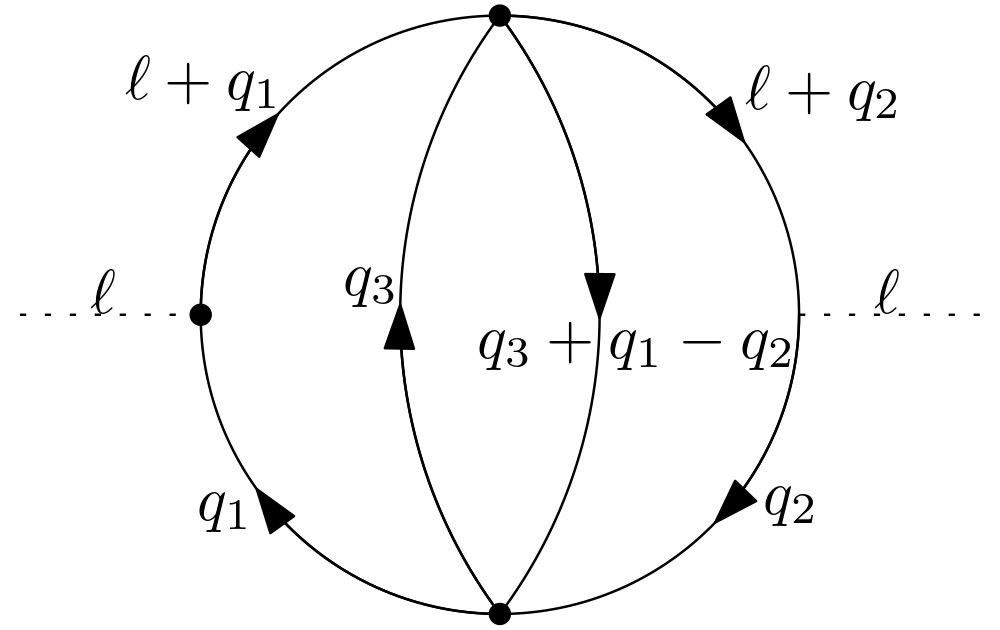}
	\caption{Eye-cat--link diagram}
	\label{fig:eyecatlink}
\end{figure}

\noindent where $I(\ell)$ is the cat's eye diagram, see fig.~\ref{fig:eyecatlink}, whose amplitude is given by
\begin{multline}
I(\ell) = \int \frac{d^D q_1}{(2\pi)^D} \int \frac{d^D q_2}{(2\pi)^D} \int \frac{d^D q_3}{(2\pi)^D} 
\frac{1}{(q_1+\ell)^2} \frac{1}{q_1^2}\times\\ \frac{1}{(q_2+\ell)^2} \frac{1}{q_2^2}
\frac{1}{(q_3+q_1-q_2)^2} \frac{1}{q_3^2}.
\end{multline}

After some manipulations, see app.~\ref{app:parametrization:eyecat}, we can obtain that the solution of cat's eye diagram at $D=4-2\varepsilon$ is
\begin{subequations}
\begin{equation}
I(\ell) = 
\frac{\Gamma\left[3\varepsilon\right]}{(4\pi)^{6-3\varepsilon} \ell^{6\varepsilon}} 
\Bigg(- c_1 +c_2 
+ \frac{1}{8\varepsilon^2} + \mathcal{O}(\varepsilon)
\Bigg),
\end{equation}
with
\begin{align}
c_1 &= 0.822\,467\,029\,8\ldots\\
c_2 &= 0.822\,467\,033\,4\ldots
\end{align}
\end{subequations}

\noindent Or, after applying a minimal subtraction scheme and keeping only the finite components
\begin{subequations}
\begin{equation}
\widehat{I}(\ell) = a_0 + a_1 \ln \ell^2 + a_2 \ln^2 \ell^2 + a_3 \ln^3 \ell^2,
\end{equation}
\noindent with constants
\begin{align}
a_3 &= -\frac{3}{4^8 \pi ^6} = -4.761\,481\,354 \times 10^{-8},\\
a_2 &= -\frac{3\gamma}{4^8 \pi^6}= -2.748\,401\,625\times 10^{-7},\\
a_1 &= 5.886\,714\,656 \times 10^{-9},\\
a_0 &= 7.074\,471\,170\times 10^{-7}.
\end{align}
\end{subequations}

\noindent We can already expect that this will generate a Gevrey-3 series due to the $\ln^3$ behavior.

\begin{figure}
	\centering
	\includegraphics[width=0.5\linewidth]{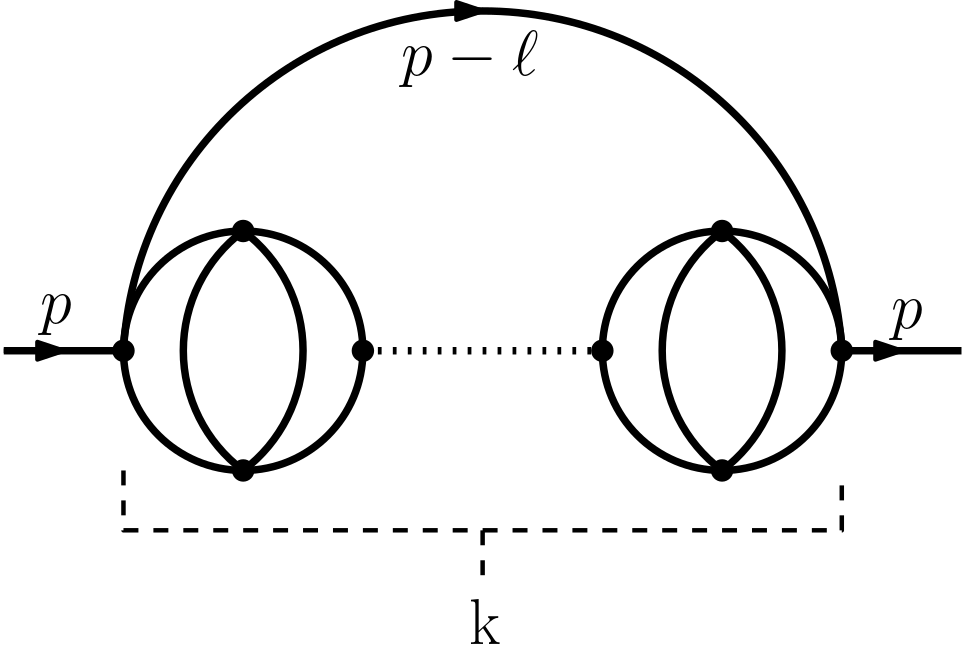}
	\caption{Chain of cat's eye organized as a correction to the mass.}
	\label{fig:eyecatchaink}
\end{figure}

We produce the chain diagram by adding the chain-links. At fig.~\ref{fig:eyecatchaink} we show the scenario with $k$ chain links. The amplitude, before regularization, is

\begin{equation}
R_k (p) = -g S_k \int \frac{d^D \ell}{(2\pi)^D} \frac{(-g^3 {I}(\ell))^k}{(p-\ell)^2},
\end{equation}
\noindent where $S_k = \left[(k+1)2^{k}\right]^{-1}$ is the symmetry factor for the diagram with $k$ eye-cat--links. We calculate the symmetry factor using a general expression for symmetry factors of the scalar theory, by ref.~\cite{Palmer:2001vq}.

At $D=4$ the function $I$ behaves asymptotically as a logarithm and by power counting $R_k$ has a quadratic divergence. We can make a subtraction as a BPHZ procedure~\cite{Itzykson:1980rh} ($R_k(p) - R_k(0) - p^2 \partial_{p^2}R_k(p)|_0$), which produces at low $p$
\begin{equation}
\widehat{R}_k (p) \sim -\frac{gp^4}{8\pi^2} \frac{1}{k+1}\left(-\frac{1}{2}g^3\right)^k \int_0^\infty \frac{d\ell}{\ell^3}\; \widehat{I}^k(\ell).
\label{eq:Rk_regularized}
\end{equation}

To investigate the asymptotic UV behavior, we can consider $\ell \in[1,\infty[$ and make the change of variables $\ell = e^t$ such that,
\begin{multline}
\widehat{R}_k (p) \sim -\frac{gp^4}{8\pi^2} \frac{1}{k+1}\left(-\frac{1}{2}g^3\right)^k \times \\\int_0^\infty dt\; e^{-2t} (a_0 + 2 a_1 t + 4 a_2 t^2 + 8 a_3 t^3)^k.
\end{multline}

\noindent As $k\in \mathbb{N}$ we can use the multinomial representation and integrate over $t$,
\begin{multline}
\widehat{R}_k (p) \sim -\frac{gp^4}{16\pi^2} \frac{1}{k+1}\left(-\frac{1}{2}g^3\right)^k \times\\
\sum_{j_0+j_1+j_2+j_3=k} a_0^{j_0}a_1^{j_1}a_2^{j_2}a_3^{j_3} 
\frac{k! \Gamma\left(1+j_1+2j_2+3j_3\right)}{j_0!j_1!j_2!j_3!}.\label{eq:Rk_js}
\end{multline}

The dominant contribution comes from $\mathbf{j}=(0,0,0,k)$ (accuracy of ${\approx70}\%$, it can be checked numerically for large values of $k$). The second contribution comes from $\mathbf{j}=(0,0,1,k-1)$ (with an accuracy of more than $90\%$). That way, we can write

\begin{equation}
\widehat{R}_k (p) \sim -\frac{gp^4}{16\pi^2} \left(1+ \frac{a_2}{3a_3} 
\right) \left(-\frac{g^3a_3}{2}\right)^k
\frac{\Gamma\left(1+3k\right)}{k+1}, \label{eq:Rk_asymp}
\end{equation}
\noindent and it is clear that these coefficients produce a divergent series. Taking the Stirling approximation for large values of $k$ the sum over all cat's eyes chains produce

\begin{subequations}
\begin{equation}
R \sim \sum_{k=1}^\infty \widehat{R}_k(p) \sim 
\kappa
\sum_{k=1}^\infty \left(\frac{\widetilde{g}}{e^3}\right)^{k}
k^{3k-\frac{1}{2}},
\end{equation}
\noindent with $\kappa$ and $\widetilde{g}$ given by
\begin{align}
\kappa &= -\sqrt{6\pi}  \frac{gp^4}{16\pi^2} \left(1+ \frac{a_2}{3a_3}\right),\\
\widetilde{g} &= -\frac{3^3}{2} g^3a_3.
\end{align}
\end{subequations}

\noindent This expression for $R$ is not summable in the first Borel plane. The Borel transform introduces a $k!$ damping that behaves just as $k^k$. We can control the divergent behavior only with a 3-Borel transform, that introduces a $k!^3$, meaning a $k^{3k}$ damping, sufficient to obtain a summable expression. Therefore, this is a Gevrey-3 asymptotic series.

If we transport eq.~\eqref{eq:Rk_asymp} into the 3-Borel plane we obtain a hypergeometric function,
\begin{multline}
B_3[R] 
\sim -\frac{gp^4}{16\pi^2} \left(1+ \frac{a_2}{3a_3} \right)
\Bigg[
-1
+\,
{}_2F_1\left(\frac{1}{3},\frac{2}{3};2;\widetilde{g} u\right)
\Bigg],
\end{multline}
\noindent it has a branch point at $u = 1/\widetilde{g}$. This cut lies in the integration path of the 3-Borel plane, therefore we have a singularity that blocks the inverse Borel transform. 

If we try to get back into the physical plane we obtain an imaginary ambiguity due to the choice of the integration path from the 3-Borel to the 2-Borel plane. However, there are no new poles in the 2-Borel plane, neither in the 1-Borel plane. Meaning that the only ambiguity appears when passing the Gevrey-3 series to the 2-Borel plane. 

The fact that the only singularity comes from the 3-Borel plane is a consequence of the previous approximation to consider the most dominant divergent behavior. In the full scenario, one would expect renormalon singularities at Gevrey-1,2 and 3. Note, for example, that at eq.~\eqref{eq:Rk_js} one of the contributions is $j_0=j_1=j_3=0,j_2=k$ which produces a Gevrey-2 series,
\begin{equation*}
\sum_{k=1}^\infty\frac{(-g^3a_2/2)^{k}}{k+1}\Gamma(1+2k),
\end{equation*}
\noindent This new series has a branch point in the 2-Borel plane located at $u = -1/(2a_2g^3)<0$, which does not lie in the integration path. If we keep investigating, we find other singularities produced by eq.~\eqref{eq:Rk_js}. In fact, the sum over the cat's eyes chain diagrams produces more then one asymptotic series, each with a different Gevrey order. However, the relevant behavior is already revealed: the highest degree is Gevrey-3 and there is a renormalon singularity that lives in the 3-Borel plane.

\section{Conclusion}

We exhibited that the renormalon problem goes beyond the first Borel plane. It seems that the renormalon singularities are in some different category when compared to the instanton-like singularities, only known to occur at the Gevrey-1 level. Although we did not discuss a realistic theory nor considered the phenomenological influence, it is beyond doubt that we have a new singularity that lives in the 3-Borel plane. We remark that our report poses a new difficulty for the program to cure all renormalon ambiguities. 
Furthermore, there are some aspects open to future investigations. We need to determine how the Gevrey-$q$ renormalon impacts the asymptotic series or the resurgence program. Also, we must check whether a Gevrey-$d$ series (with $d>1$) also appear for realistic theories.

\acknowledgments{The author thanks the Brazilian agency \textit{Conselho Nacional de Desenvolvimento Cient\'ifico e Tecnol\'ogico} (CNPq) for financial support.}

\appendix


\section{Cat's eye diagram}
\label{app:parametrization:eyecat}

In this section, we consider for completeness a step-by-step computation of the cat's eye diagram, as this evaluation is not easily found in textbooks.

Let us start from the original amplitude,

\begin{multline}
I(\ell) = \int \frac{d^D q_1}{(2\pi)^D} \int \frac{d^D q_2}{(2\pi)^D} \int \frac{d^D q_3}{(2\pi)^D} 
\frac{1}{(q_1+\ell)^2} \frac{1}{q_1^2}\times\\ \frac{1}{(q_2+\ell)^2} \frac{1}{q_2^2}
\frac{1}{(q_3+q_1-q_2)^2} \frac{1}{q_3^2}, 
\end{multline}
\noindent and then introduce Schwinger parameters $\alpha_i$ for each propagator,
\begin{multline}
I(\ell) = 
\int \frac{d^D q_1}{(2\pi)^D} 
\int \frac{d^D q_2}{(2\pi)^D} 
\int \frac{d^D q_3}{(2\pi)^D} 
\int_0^\infty \prod_{i=1}^6 d\alpha_i\\
e^{- \alpha_1 (q_1+\ell)^2 -\alpha_2 q_1^2}
e^{- \alpha_3 (q_2+\ell)^2 -\alpha_4 q_2^2}
e^{- \alpha_5 (q_3+q_1-q_2)^2 -\alpha_6 q_3^2}.
\end{multline}
After some algebraic manipulations on these expressions, we obtain that
\begin{multline}
I(\ell) = 
\int \frac{d^D q_1}{(2\pi)^D} 
\int \frac{d^D q_2}{(2\pi)^D} 
\int \frac{d^D q_3}{(2\pi)^D} 
\int_0^\infty \prod_{i=1}^6 d\alpha_i\\
e^{-\alpha_{125}\left(q_1 + \frac{\ell \alpha_1 + q_3 \alpha_5}{\alpha_{125}}\right)^2}
e^{- \alpha_{346} \left(q_2 + \frac{\ell \alpha_3 + q_3 \alpha_6}{\alpha_{346}}\right)^2}
\times\\
e^{- \frac{U}{\alpha_{125}\alpha_{346}}
	\left[
	q_3 
	- \ell \frac{\alpha_1\alpha_5 \alpha_{346}+\alpha_3\alpha_6 \alpha_{125}}{U}
	\right]^2}
e^{- \frac{V \ell^2}{U}},
\end{multline}
\noindent where
\begin{subequations}
\begin{align}
\alpha_{i_1 i_2 \ldots i_k} &= \alpha_{i_1} + \alpha_{i_2} + \cdots + \alpha_{i_k},\\
U &= \alpha_{12}\alpha_5 \alpha_{346} + \alpha_{34}\alpha_{6}\alpha_{125},\\
V &= \alpha_1 \alpha_2 (\alpha_{34} \alpha_{56} + \alpha_5 \alpha_6)+ \alpha_3 \alpha_4 (\alpha_{12}\alpha_{56} + \alpha_5 \alpha_6)\nonumber\\
&
+ \alpha_5 \alpha_6 (\alpha_1 \alpha_4 + \alpha_2 \alpha_3).
\end{align}
\end{subequations}

We can make a shift in the momenta,
\begin{multline}
I(\ell) = 
\int \frac{d^D q_1}{(2\pi)^D} 
\int \frac{d^D q_2}{(2\pi)^D} 
\int \frac{d^D q_3}{(2\pi)^D} 
\int_0^\infty \prod_{i=1}^6 d\alpha_i\\
e^{-\alpha_{125}q_1^2}
e^{- \alpha_{346} q_2^2}
e^{- \frac{U}{\alpha_{125}\alpha_{346}} q_3^2}
e^{- \frac{V \ell^2}{U}},
\end{multline}
\noindent and then compute the integration of the internal momenta

\begin{equation}
I(\ell) = 
\frac{1}{(4\pi)^{\frac{3D}{2}}} 
\int_0^\infty \prod_{i=1}^6 d\alpha_i
\frac{e^{- \ell^2 \frac{V}{U}}}{U^{\frac{D}{2}}}.
\label{eq:app:I}
\end{equation}

Let us reparametrize defining the new sectors
\begin{align*}
\alpha_1 &= s t_1 \cdots t_5,\\
\alpha_2 &= s t_1 \cdots t_4 (1-t_5),\\
\alpha_3 &= s t_1 t_2 t_3 (1-t_4),\\
\alpha_4 &= s t_1 t_2 (1-t_3),\\
\alpha_5 &= s t_1 (1-t_2),\\
\alpha_6 &= s (1-t_1),
\end{align*}
\noindent such that $\alpha_1+\ldots+\alpha_6 = s$. This change of variables produce the Jacobian
\begin{equation*}
d\alpha_1 \cdots d\alpha_6 = s^5 t_1^4 t_2^3 t_3^2 t_4 ds dt_1 \cdots dt_5,
\end{equation*}
\noindent and after some manipulation we obtain that
\begin{subequations}
\begin{align}
\alpha_{12} &= s t_1 \cdots t_4,\\
\alpha_{34} &= s t_1 t_2 (1-t_3 t_4),\\
\alpha_{125} &= s t_1 (1-t_2 +  t_2t_3t_4),\\
\alpha_{346} &= s \left(1-t_1+t_1t_2(1-t_3t_4)\right),\\
U &= s^3 t_1^2 t_2 \overline{U}, \\
V &= s^4 t_1^3 t_2^2 t_3 \overline{V},
\end{align}
\end{subequations}
\noindent with $\overline{U}$ and $\overline{V}$ given by
\begin{multline}
\overline{U} = (1-t_2) t_3 t_4 (1-t_1+t_1t_2(1-t_3t_4))
\\+ (1-t_2+t_2t_3t_4)(1-t_3t_4)(1-t_1),
\end{multline}
\begin{multline}
\overline{V} = 
(1-t_1)(1-t_2)(1-t_4+t_4t_5) (1-t_3+t_3t_4 (1-t_5))\\
+t_2t_3t_4 (1-t_1t_2)
\left(
(1-t_3)(1-t_4) + (1-t_3t_4)(1-t_5)t_4t_5
\right).
\end{multline}
Substituting this back into the integral $I$, eq.~\eqref{eq:app:I}, we obtain

\begin{multline}
I(\ell) = 
\frac{1}{(4\pi)^{\frac{3D}{2}}} 
\int_0^1 \prod_{i=1}^5 dt_i
\frac{t_1^{4-D} t_2^{3-\frac{D}{2}} t_3^2 t_4}{\overline{U}^{\frac{D}{2}}(\boldsymbol{t}) }
\\\int_0^\infty ds\;
s^{\left(6-\frac{3D}{2}\right)-1}
e^{- s t_1t_2t_3\ell^2 \frac{\overline{V}}{\overline{U}}}.
\end{multline}

\noindent This expression can be directly integrated over $s$,

\begin{equation}
I(\ell) = 
\frac{\Gamma\left[6-\frac{3D}{2}\right]}{(4\pi)^{\frac{3D}{2}}} 
\int_0^1 \prod_{i=1}^5 dt_i
\frac{t_1^{-2+\frac{D}{2}} t_2^{-3+D} t_3^{-4+\frac{3D}{2}} t_4}{\overline{U}^{2D-6} \overline{V}^{6-\frac{3D}{2}} \ell^{12-3D}},
\end{equation}
\noindent and taking that $D=4-2\varepsilon$,
\begin{equation}
I(\ell) = 
\frac{\Gamma\left[3\varepsilon\right]}{(4\pi)^{6-3\varepsilon} \ell^{6\varepsilon}} 
\int_0^1 \prod_{i=1}^5 dt_i
\frac{t_1^{-\varepsilon} t_2^{1-2\varepsilon} t_3^{2-3\varepsilon} t_4}{\overline{U}^{2-4\varepsilon} \overline{V}^{3\varepsilon}}.
\label{eq:app:I2}
\end{equation}

Although at this point we do not know the behavior of the remaining integral with respect to $\varepsilon$, we can already see that the pure logarithmic behavior of $\ell$ can \textit{only} occur at $\varepsilon=0$, which reinforces the perception that the existence of renormalons is an aspect related to the dimension where the theory is renormalizable.

Let us take $t_1^{-\varepsilon} \overline{V}^{-3\varepsilon}$ as a subdominant contribution ($1+\mathcal{O}(\varepsilon)$). And as the polynomial $\overline{U}$ is independent of $t_5$ this variable can be integrated out. Also we can rewrite $\overline{U}$ as
\begin{multline}
\overline{U} = 
(1-t_2+t_2t_3t_4 (1-t_3t_4) )
\\- t_1 (1 - t_2 + t_2^2 t_3 t_4(1-t_3t_4))= A_1 -t_1 A_2,
\end{multline}
\noindent to integrate over $t_1$
\begin{multline}
\int_0^1 dt_1 \frac{1}{(A_1 - t_1 A_2)^{2-4\varepsilon}} = \int_{A_1}^{A_1-A_2} \frac{dy}{-A_2} \frac{1}{y^{2-4\varepsilon}} \\= \frac{1}{1-4\varepsilon}
\left[
\frac{1}{(A_1-A_2)^{1-4\varepsilon}A_2} - \frac{1}{A_1^{1-4\varepsilon}A_2}
\right].
\end{multline}

At this point, eq.~\eqref{eq:app:I2} becames
\begin{equation}
I(\ell) = 
\frac{\Gamma\left[3\varepsilon\right](1+\mathcal{O}(\varepsilon))}{(4\pi)^{6-3\varepsilon} \ell^{6\varepsilon}} 
J_1
-
\frac{\Gamma\left[3\varepsilon\right](1+\mathcal{O}(\varepsilon))}{(4\pi)^{6-3\varepsilon} \ell^{6\varepsilon}} 
J_2,
\end{equation}
\noindent where $J_1$ and $J_2$ are given by
\begin{multline}
J_1 = \int_0^1 dt_2 dt_3 dt_4 
\frac{t_2^{1-2\varepsilon} t_3^{2-3\varepsilon} t_4}{1 - t_2 + t_2^2 t_3 t_4(1-t_3t_4)}\times \\
\frac{1}{\left[t_2t_3t_4(1-t_2)(1-t_3t_4)\right]^{1-4\varepsilon}},
\end{multline}
\begin{multline}
J_2 = \int_0^1 dt_2 dt_3 dt_4
\frac{t_2^{1-2\varepsilon} t_3^{2-3\varepsilon} t_4}{1 - t_2 + t_2^2 t_3 t_4(1-t_3t_4)}\times \\
\frac{1}{\left[1-t_2+t_2t_3t_4 (1-t_3t_4)\right]^{1-4\varepsilon}}.
\end{multline}
The second integral, $J_2$, is finite and gives
\begin{equation}
J_2 = 0.822\,467\,029\,8\ldots = c_1.
\end{equation}

The integral $J_1$ has a divergent contribution from the $t_2$ integration. We can reorganize it and compute the other integrals 
\begin{multline*}
J_1 
= 
\int_0^1 dt_2 
\frac{t_2^{2\varepsilon}}{(1-t_2)^{1-4\varepsilon}}
\int_0^1 dt_3 \int_0^1 dt_4\times\\
\frac{1}{1 - t_2 + t_2^2 t_3 t_4(1-t_3t_4)}
\frac{t_3}{(1-t_3t_4)^{1-4\varepsilon}}\\
= 
\int_0^1 dt\,
t^{2\varepsilon-1} (1-t)^{4\varepsilon-1}
\frac{2 \ln(1-t)}{t-2}.
\end{multline*}

Note that we can rewrite it as a derivative of a function of $\mu$ and $\nu$ (and later identify $\mu=2\varepsilon$, $\nu=4\varepsilon$)
\begin{align}
G(\mu,\nu) = - \int_0^1 dt\,
t^{\mu-1} (1-t)^{\nu-1} \left(1-\frac{t}{2}\right)^{-1},
\\
\partial_\nu G(\mu,\nu) = \int_0^1 dt\,
t^{\mu-1} (1-t)^{\nu-1}
\frac{2 \ln(1-t)}{t-2}.
\end{align}

Where is $G(\mu,\nu)$ is known, see eq.~3.197(3) from ref.~\cite{Gradshteyn:7th},
\begin{align}
G(\mu,\nu) &= - B(\mu,\nu) {}_2F_1\left(1,\mu;\mu+\nu;\frac{1}{2}\right) \nonumber\\&= - \sum_{k=0}^\infty \frac{\Gamma(\mu+k)\Gamma(\nu)}{\Gamma(\mu+\nu+k)} \frac{1}{2^k},
\end{align}
\noindent where $B(\mu,\nu)$ is the Beta function and ${}_2F_1$ is the hypergeometric function. We use the representation as a infinite sum of gamma functions, to make manipulations easier.

Making the derivative with respect to $\nu$ and substituting $\mu=2\varepsilon, \nu=4\varepsilon$,
\begin{multline}
J_1 = G^{(0,1)}(2\varepsilon,4\varepsilon) =
- \sum_{k=0}^\infty \frac{\Gamma(2\varepsilon+k)}{2^k \Gamma(6\varepsilon+k)}
\Bigg(
\Gamma'(4\varepsilon)\\
-
\frac{\Gamma(4\varepsilon)\Gamma'(6\varepsilon+k)}{\Gamma(6\varepsilon+k)}
\Bigg)
=
\left( \frac{1}{16\varepsilon^2} + \frac{\pi^2}{6} 
+ \mathcal{O}(\varepsilon)
\right)
\\+\Bigg[
\left( \frac{1}{16\varepsilon^2} - \frac{\pi^2}{12} - \frac{\gamma^2}{2}
\right)
\sum_{k=1}^\infty \frac{1}{2^k}
- \gamma \sum_{k=1}^\infty \frac{\psi^{(0)}(k)}{2^k}
\\
- \frac{1}{2} \sum_{k=1}^\infty \frac{\left(\psi^{(0)}(k)\right)^2}{2^k}
+ \frac{1}{2} \sum_{k=1}^\infty \frac{\psi^{(1)}(k)}{2^k}
+ \mathcal{O}(\varepsilon)\Bigg]=\\
\frac{1}{8\varepsilon^2}
+\frac{\pi^2}{8}
+\frac{\gamma^2}{2}-\gamma \ln 2 +\frac{1}{4}\ln^2 2 
- \frac{1}{2} \sum_{k=1}^\infty \frac{\left(\psi^{(0)}(k)\right)^2}{2^k}+ \mathcal{O}(\varepsilon)=\\
\frac{1}{8\varepsilon^2} + \underbrace{0.822\,467\,033\,4\ldots}_{c_2}+ \mathcal{O}(\varepsilon) 
\end{multline}

Finally, we obtain 

\begin{equation}
I(\ell) = 
\frac{\Gamma\left[3\varepsilon\right]}{(4\pi)^{6-3\varepsilon} \ell^{6\varepsilon}} 
\Bigg(- c_1 +c_2 
+ \frac{1}{8\varepsilon^2} + \mathcal{O}(\varepsilon)
\Bigg).
\end{equation}

We can expand with respect to $\varepsilon\rightarrow0$ to make the poles evident,
\begin{multline}
I(\ell) = \frac{1}{(4\pi)^6}
\left(\frac{1}{3\varepsilon}-\gamma+\frac{\varepsilon}{4} \left(6\gamma^2+\pi^2\right) + \mathcal{O}(\varepsilon^2)\right)\\
\left(1 - 3\varepsilon \ln \frac{\ell^2}{4\pi} +
\frac{9}{2}\varepsilon^2 \ln^2 \frac{\ell^2}{4\pi}
-\frac{9}{2} \varepsilon^3 \ln^3 \frac{\ell^2}{4\pi}
+\mathcal{O}(\varepsilon^4)
\right)\times\\
\Bigg(\frac{1}{8\varepsilon^2}  - c_1 +c_2
+ \mathcal{O}(\varepsilon)
\Bigg).
\label{eq:epsilonexpansion}
\end{multline}

After some algebraic manipulations it produces
\begin{multline}
I(\ell) = \frac{1}{(4\pi)^6} \Bigg[
\frac{1}{8\varepsilon^3}
+ \frac{1}{\varepsilon^2} 
\left(-\frac{\gamma}{8}-\frac{1}{8}\ln\frac{\ell^2}{4\pi}\right)\\
+ \frac{1}{\varepsilon} 
\left(
\frac{c_2-c_1}{3}+\frac{\pi^2+6\gamma^2}{32}
-\frac{9\gamma}{16}\ln\frac{\ell^2}{4\pi}
+\frac{3}{16}\ln^2\frac{\ell^2}{4\pi}
\right)\\
- 3 \ln \frac{\ell^2}{4\pi} 
\left(\frac{c_2-c_1}{3}+\frac{\pi^2+6\gamma^2}{32}\right)
\\- \frac{3\gamma}{16} \ln^2 \frac{\ell^2}{4\pi}
- \frac{3}{16}\ln^3 \frac{\ell^2}{4\pi}
+ a_0 + \mathcal{O}(\varepsilon)
\Bigg]
\end{multline}

We consider a $\overline{MS}$ subtraction, such that the poles ($\varepsilon^{-3},\varepsilon^{-2},\varepsilon^{-1}$) are removed and the amplitude becames
\begin{subequations}
\begin{equation}
\widehat{I}(\ell) = a_0 + a_1 \ln \ell^2 + a_2 \ln^2 \ell^2 + a_3 \ln^3 \ell^2,
\end{equation}
\noindent where the constants are
\begin{align}
a_3 = -\frac{3}{4^8 \pi ^6} = -4.761\,481\,354 \times 10^{-8},\\
a_2 = -\frac{3\gamma}{4^8 \pi^6}= -2.748\,401\,625\times 10^{-7},\\
a_1 = 5.886\,714\,656 \times 10^{-9}.
\end{align}
\end{subequations}

The correct determination of $a_0$ requires one more term in the $\varepsilon$-expansion in eq.~\eqref{eq:epsilonexpansion}. After some computations one finds $a_0 = 7.074\,471\,170\times 10^{-7}.$

Therefore, the chain of cat's eyes has the $\ln^3 \ell$ behavior that can produce a Gevrey-3 divergent series. This logarithmic behavior is a property of $D = 4$ and is responsible for the renormalon divergence. If we consider the same diagram in any other dimension we will obtain something like $\ell^a \ln^b \ell$ that does not produce a renormalon divergence. 


\bibliography{Refs}{}
\bibliographystyle{apsrev4-1}

\end{document}